# SIX-Trust for 6G: Towards a Secure and Trustworthy 6G Network

Yiying Wang, Xin Kang, *Senior Member, IEEE,* Tieyan Li, *Member, IEEE,* Haiguang Wang, *Senior Member, IEEE,* Cheng-Kang Chu, *Member, IEEE,* and Zhongding Lei, *Senior Member, IEEE*

*Abstract*— Recent years have witnessed a digital explosion with the deployment of 5G and proliferation of 5G-enabled innovations. Compared with 5G, 6G is envisioned to achieve much higher performance in terms of latency, data rate, connectivity, energy efficiency, coverage and mobility. To fulfil these expectations, 6G will experience a number of paradigm shifts, such as exploiting new spectrum, applying ubiquitous ML/AI technologies and building a space-air-ground-sea integrated network. However, these paradigm shifts may lead to numerous new security and privacy issues, which traditional security measures may not be able to deal with. To tackle these issues and build a trustworthy 6G network, we introduce a novel trust framework named as SIX-Trust, which composes of 3 layers: sustainable trust (S-Trust), infrastructure trust (I-Trust) and xenogenesis trust (X-Trust). Each layer plays a different role, and the importance of each layer varies for different application scenarios of 6G. For each layer, we briefly introduce its related enabling technologies, and demonstrate how these technologies can be applied to enhance trust and security of the 6G network. In general, SIX-Trust provides a holistic framework for defining and modeling trust of 6G, which can facilitate establishing a trustworthy 6G network.

## I. INTRODUCTION

THE rapid development of 5G has opened up a world of low-latency communication, high-speed data delivery, and exponentially increased connectivity among numerous devices and sensors. Deployment of network function virtualization (NFV) and virtual network functions (VNF) in 5G offers a dynamic network architecture and enables flexible resource allocation by separating service from hardware. Due to unprecedented growth of data volume and expanding demand for ubiquitous connectivity, researchers have been driven to put much effort on developing 6G technologies. It is expected that 6G needs to offer at least 20 times more network capacity, and 50 times more data transmission rate than 5G. Besides, 6G is envisioned to add one more dimension to create a three-dimensional network covering from terrestrial to non-terrestrial, from space to underwater, and form the so-called space-air-ground-sea integrated network. Moreover, as an essential enabler, Artificial Intelligence (AI) empowers automation of network element creation and operation, anomaly detection and ubiquitous system monitoring. Proliferation of pervasive intelligence will drive the transformation of mobile communications from "connected everything" to "connected intelligence" [1].

However, the expanding connectivity and new features of 6G may introduce new security threats from multiple aspects: open interfaces, pervasive usage of NFV and VNF, the integration of sensing and computing, extensive usage of cloud and edges, the complicated relationship among human, things, and connected intelligence, and the entangled relationship among various involved stakeholders. Millions of connected devices and sensors, forming the base layer of threats, will increase the vulnerability to the impact of attacks. More threats are likely to appear with the extensive use of AL/ML. For instance, the training procedure of AI models requires substantial amount of data, which may lead to data leakage or malicious usage of sensitive and confidential information.

To deal with all aforementioned potential threats, it is believed that 6G needs to be designed as a trustworthy network in nature. Trust needs to be evaluated across end-devices, access networks, and core networks. However, trust is a very complicated concept and still under-studied in both academia and industry. Thus, how to build a trustworthy 6G network becomes a challenging research problem. In this paper, we present our points of view on how to build a trustworthy 6G network. We adopt a systematic approach to investigate and analyze trust challenges of 6G in a layer-wise manner, and we propose a three-layer novel trust framework for 6G to assure the 6G network trustworthiness. The proposed trust framework, consisting of Sustainable trust (S-Trust), Infrastructure trust (I-Trust) and Xenogenesis trust (X-Trust), is named as SIX-Trust. Each layer plays a different role, and the importance of each layer varies for different application scenarios of 6G. For each layer, we share our views on relevant technologies as potential solutions to address the security and trust challenges. The proposed framework is envisioned to provide a comprehensive overview of 6G trust and offer insights into viable methods that can be applied to assure 6G network's trustworthiness.

## II. SIX-TRUST FOR 6G

From our perspective, to build a trustworthy 6G network, we need a multi-layer hierarchical trust architecture, consisting of a root layer, a foundation layer and a representation layer. Based on these, we propose the three-layer SIX-Trust, comprising of S-Trust, I-Trust and X-Trust.

### A. Sustainable Trust (S-Trust)

Sustainable trust (S-Trust) embodies trust representation, which measures the degree of trust of users. In the context of 6G, continual and tenable trust representation and evaluation are indispensable for sustainably providing 'a sense of trust'. The sustainable trust is most palpable layer of all layers, as trust is often presented in numerical values or visual artifacts.

To increase perceivability, trust needs to be quantified and sustainably assessed by both static and dynamic evaluation processes, which may be facilitated by the AI technology. Thus, with the capability to preserve data privacy, achieve better explainability, and mitigate undesirable biases, trustworthy AI becomes a critical enabler for sustainable trust. Moreover, trust evaluation will foster establishment of trust relationships, as it becomes more convenient for evolved parties to have a sense of other parties' trustworthiness. Decentralized trust and federated trust are two representatives envisioned to be of great importance to 6G for achieving S-Trust.

*B. Infrastructure Trust (I-Trust)*

Infrastructure trust (I-Trust) reinforces the trust from the bottom layer and provides a trustworthy architecture for the upper layer. Given countless security threats posed by involvement of various types of entities and inclusion of all kinds of cutting-edge technologies, a reliable and trustworthy network infrastructure is thus crucial for underpinning secure 6G networks by assuring secure execution of upper layer applications. I-Trust reflects the trustworthiness of network architecture, which is envisioned to be distributed and autonomous. The architecture is undergirded by trustworthy underlay such as DPKI and NFVI, which can be viewed as skeleton of I-Trust, for providing decentralized authentication topology and facilitating more flexible deployment of intelligence by virtualization. On top of the underlay, trustworthy protocols further reinforce the trustworthiness during communication.

*C. Xenogenesis Trust (X-Trust)*

Xenogenesis trust (X-Trust) represents the foundation of trust, the initial point where the chain of trust starts in 6G networks. It is an endogenous trustworthiness originated from three aspects: trusted foundation, trusted platform, as well as trusted hardware. The involved technologies are inherently designed to be trustworthy in order to preserve network security. For instance, TPM as an embedded security hardware, is inherently resistant to spoofing and tampering. It is therefore can be used as the basis for hardware root of trust, by providing hard-level protection. In other words, X-Trust stems from the beliefs about in-built security features of the related technology. It is a kind of technology trust. Consequently, X-Trust offers higher trust degree. It also forms the basis of SIX-Trust.

*D. Relationship among Three Layers*

Three layers are stacked in a hierarchical manner, each of which possesses different degree of trust. This three-layer model is inspired by the model of earth. As shown in Fig.1, X-Trust, which is born with trust, has the highest degree of trust and is the core of SIX-Trust. I-Trust, like the earth's mantle, is mainly responsible for reinforcing X-Trust and supporting the S-Trust. S-Trust, like the earth's crust, is more perceptible and directly faces the network users. These three-layers are tightly connected and influence one another; however, it is worth pointing out that the importance of each layer varies depending on the type of networks. For example, IoT includes numerous devices and sensors with limited processing ability and insufficient security mechanisms [4], which significantly undermines the X-Trust. Moreover, the degree of S-Trust and I-Trust in IoT are both low and weak due to resource limitation. On the contrary, cellular networks are expected to achieve high degrees of trust of all three layers. First, advanced encryption methods and trusted algorithms will enable a strong X-Trust. Moreover, more trustworthy and secure authentication methods will be applied in order to ensure a solid I-Trust. Last but not least, from the users' perspective, the degree of trust representation and trust perceivability will be greatly increased, since cellular networks will shift from being device-centric to becoming user-centric [5]. The importance of each layer for IP networks and NTT are illustrated in Fig 1, and will not be discussed here in details due to the page limit.

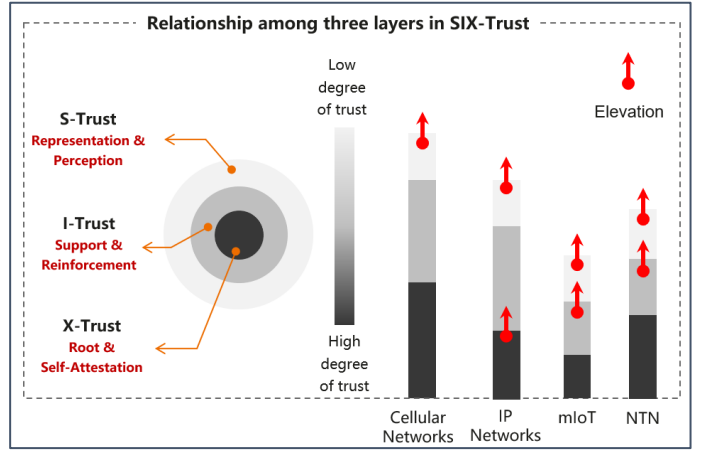

**Figure 1**. Relationship among three layers in SIX-Trust and their ratios in different 6G scenarios (Cellular Networks, IP Networks, mIoT, NTN).

## III. SUSTAINABLE TRUST

In this section, we present the potential technologies that are crucial to achieve sustainable trust.

*A. Trustworthy AI*

With the rapid development of network softwarization and virtualization, ubiquitous AI has become a trend for 6G to build self-adapting, self-sustaining and self-learning networks. The beneficial relationship between AI and 6G networks is mutual: AI empowers 6G in terms of automation, attack detection and defense, semantic communication, optimal resource management, trust evaluation, and efficient network maintenance, while 6G provides massive data and trustworthy infrastructure for AI. This relationship is also known as "Network for AI and AI for Network" [6]. In this subsection, we will discuss three key components of trustworthy AI: privacy-preserving AI, explainable AI, and unbiased AI.

**1) Privacy-preserving AI**

The training process of AI models requires massive data collected through networks, which will pose great threats to users' privacy. Federated learning (FL), being decentralized and distributed, offers a solution to privacy-preserving learning by facilitating computing on edge devices of 6G networks. FL allows on-device training, which means that data can be processed locally, and a shared model will be trained in a collaborative manner. Wireless devices with FL only need to upload their model parameters to the base station, instead of exchanging the entire training dataset. Thus, user's data will be processed in a local and distributed manner. Trustworthiness of FL can be further strengthened by robust aggregation algorithms for secure learning and differential privacy for privacy preserving. It has also been suggested that distributed ledger technology (DLT) can be converged with FL to form a distributed and trustworthy machine learning system [7].

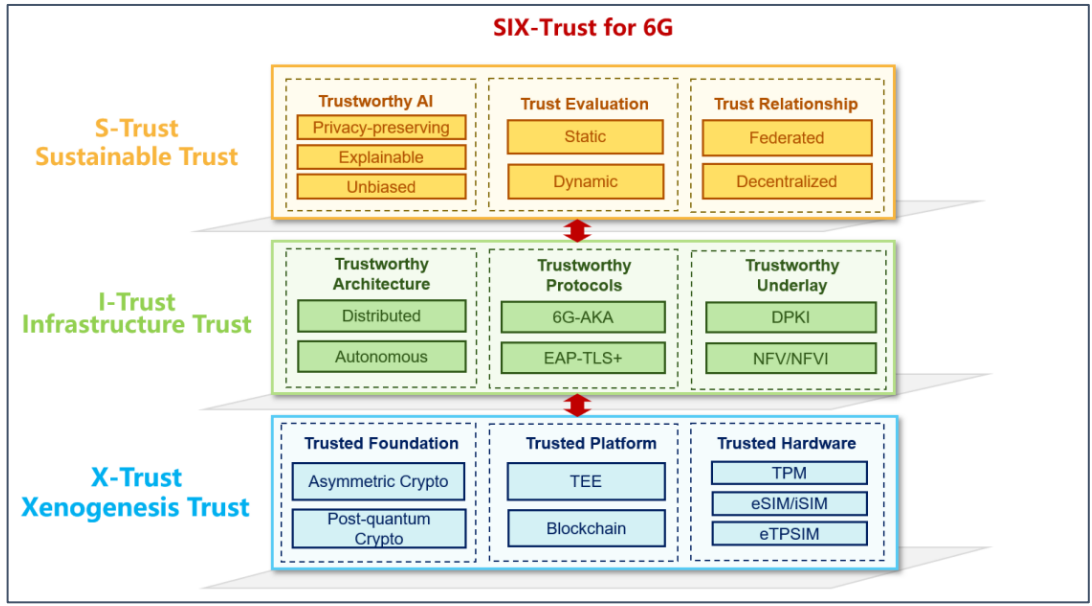

**Figure 2**. The SIX-Trust Framework

**2) Explainable AI**

In the context of 6G, understanding of decision-making process is extremely important especially in the fields which involve physical interactions between human body and machine, where safety is the primary concern. However, a number of state-of-the-art (SOTA) models of AI are often depicted as black boxes due to lack of transparency and explainability. In other words, users usually have no idea about how and why a decision is made by the model, which will greatly undermine the model's trustworthiness and make the decision inconvincible. Explainable AI (XAI) is therefore proposed in order to help human to understand the decision-making process, provide new insights of the data from an AI's perspective, and most importantly, facilitate trust establishment among AI and people. XAI is used to explain a black-box model logically or mathematically, providing a decision of trust. The explainability and transparency of AI will assist 6G stakeholders to better design their strategies for AI development and integration.

**3) Unbiased AI**

There are two major concerns related to fairness in AI models in 6G: bias and discrimination. Bias mainly originates from inappropriate data collection or flawed design of algorithms, while discrimination is often caused by stereotypes towards certain sensitive features, and can be derived from existing biases. Existence of bias in training data can lead to a biased training process, and eventually produce a biased AI model. This type of bias is likely to cause the model to be discriminative against certain attributes, which will downgrade the model's performance. Hence, assuring an unbiased dataset in the data collection and pre-processing is critical for building a fair trust evaluation model.

Biased AI models will significantly affect 6G network trustworthiness, and reduce reliability of trust evaluation process. To tackle this issue, two solutions have been proposed: fairness toolkits, which can be accessed as functions to detect and evaluate bias in a model quantitatively, and fairness checklist, which offers comprehensive guide to ensure fairness.

*B. Trust Evaluation*

To build trustworthy 6G networks, trust needs to be evaluated across numerous devices and heterogenous networks in both static and dynamic way.

**1) Static Trust Evaluation**

Static trust is often measured against network hardware. Devices have static properties which will not change over time, such as their manufacturer, their hardware firmware, their software configuration. In 6G networks, large amount of these kinds of information will be collected from devices for the purpose of static trust evaluation. Thus, static trust of the devices will have a significant impact on overall trust of the whole networks.

An example for static trust evaluation would be the Common Criteria (CC) for Information Technology Security Evaluation, which is an international standard for security evaluation. The standard evaluates reliability of a device mainly based on security functions provided by the device. Device identity, including device's certificate, vendor's information, etc., may also be used for static trust assessment.

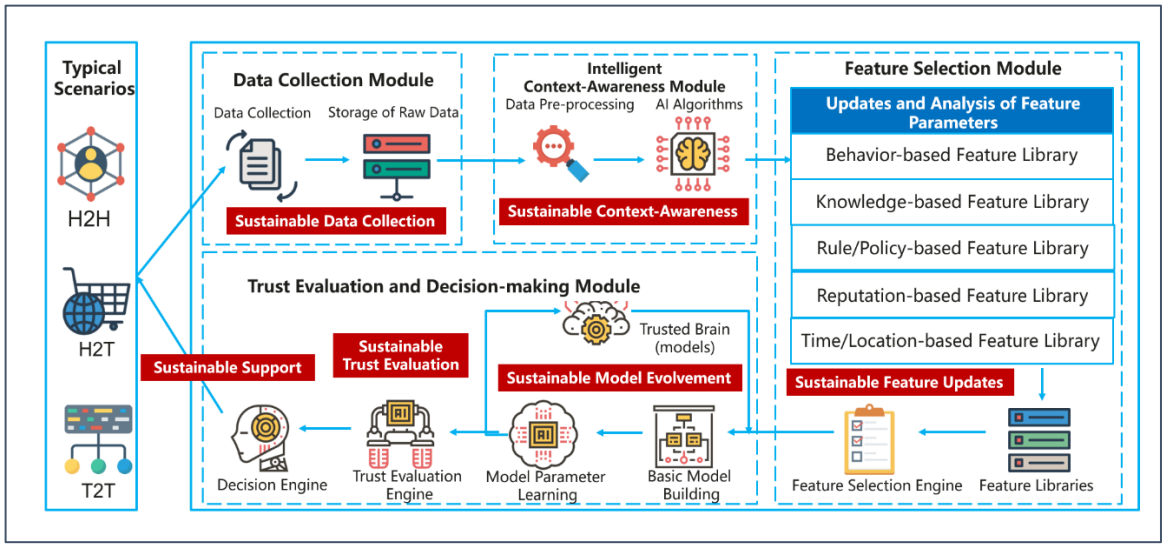

**Figure 3**. An illustration of dynamic trust evaluation framework

### 2) Dynamic Trust Evaluation

Dynamic trust evaluation is usually used for continuously monitoring users, devices, and applications' behaviors. To assure security and trustworthiness of communication in a highly connected and heterogenous environment, an integrated real-time dynamic trust evaluation framework that is applicable in diverse scenarios is essentially needed. As shown in Fig. 3, in our opinion, the dynamic trust evaluation framework can be divided into four modules that are interconnected:

**Data collection module** continually collects data generated by the evaluatee and stores raw data in database. **Intelligent context-awareness module** pre-processes stored data, and then apply AI/ML algorithms for feature extraction. **Feature selection module** contains multiple feature libraries that will be updated by previously-extracted features. Appropriate feature parameters for current scenario are selected by feature selection engine, and passed to next module to build and train AI/ML models. **Trust evaluation and decision-making module** takes in selected feature parameters and finetunes AI/ML models for trust evaluation. Decisions will be made based on trust values after evaluation, and will be used to support the policy control for different 6G scenarios.

### C. Trust Relationship

In this subsection, we discuss two promising trust relationships, decentralized trust and federated trust, for secure and seamless 6G identity management.

### 1) Decentralized Trust

In existing networks, identity management (IDM) of network devices are mainly supported by a centralized PKI where certificates are issued from a certificate authority (CA). The trust relationship between the CA and related entities is highly centralized, which may lead to many security issues, such as single point of failure. For instance, a compromised root CA will result in sharp decrease of trustworthiness of sub-CAs as well as issued certificates.

To build a more trustworthy 6G network, IDM for network devices should be structured in a decentralized manner. One promising enabler would be the decentralized identifiers (DID), which has been announced by W3C as an official Web standard. DID is not dependent on any central issuing agency (e.g., CA), and are verifiable cryptographically. The decentralized IDM is envisioned to enable key management without CAs across network slices, facilitate trustworthy and secure mutual authentication for massive IoT devices, with interoperability provided, and offer better privacy protection.

### 2) Federated Trust

As 6G networks will encompass unprecedented amount of heterogenous devices and diversified services, federated trust becomes extremely critical to network security. To address the issue of cumbersome management of numerous user credentials, identity federation has been proposed as a feasible solution to provide a more cohesive authentication process. More importantly, it establishes federated trust relationship between an identity provider (IdP) and a service provider (SP). The SPs do not need to directly handle users' credentials, but rather authenticate users based on federated trust on the IdP. Related standards include but does not limit to OAuth 2.0 and OpenID Connect, both of which can be applied to implement single sign-on (SSO) scheme. Another way to build federated trust is to form network service federation by orchestrating network services across multiple domains.

Establishing federated trust relationship will bring numerous benefits to 6G networks, one of which is that network security will be strengthened, since simplification of registration process will reduce security breaches caused by numerous user credentials and login interfaces. In addition, users will have a seamless experience across multiple domains and applications, as the operators are projected to orchestrate services of external domains. Federation also facilitates secure and effective resource sharing among different entities, increases network flexibility, as well as reduces operational cost.

## IV. INFRASTRUCTURE TRUST

In this section, we present the potential technologies that are crucial to reinforce infrastructure trust.

*A. Trustworthy architecture*

With the exponentially increasing demand in ubiquitous connectivity, traditional centralized network architectures may be no longer applicable. Network architecture of 6G is envisioned to be highly distributed and autonomous [8].

Compared to former centralized architecture, distributed and decentralized architecture will be more resilient to external threats. As 6G networks need ubiquitous and uniform coverage, cell-free MIMO has been advocated to elevate uplink capacity and avoid inter-cell interference. Cell-free massive MIMO deploys antennas in a distributed manner, where cells and cell boundaries no longer exist. It takes the advantages of both massive MIMO and distributed systems for more reliable and user-centric communication networks. 6G network architecture will also be more distributed in the form of distributed computing (e.g., edge computing), decentralized data storage, decentralized identity management, as well as decentralized AI (e.g., federated learning).

Meanwhile, it is envisioned that 6G networks will adopt an autonomous network architecture for achieving self-provisioning, self-recovering and self-evolving abilities. Autonomous networks (AN) are supposed to benefit both 6G stakeholders and users, by providing optimized resource allocation, enhanced network scalability, and increased operations and maintenance efficiency. Meanwhile, AN can facilitate flexible network deployment in diverse scenarios, which presents a solution to complex control plane of distributed architecture. It is expected that network operators will gradually entrust their control authority as well as management duties to self-sustaining AN of 6G.

*B. Trustworthy Protocols*

In this subsection, two emerging authentication protocols for ensuring secure communications are presented.

1) **6G Authentication and Key Agreement (6G-AKA)**

AKA is a security protocol specified by 3GPP which enables mutual authentication between end-user and the core network. Given the complexity of 6G network and a number of novel applications (e.g., tele-medical, tele-presence holography, tele-operation of industry machines), 6G-AKA is required to provide more fast, reliable, and trustworthy authentication. As a precedent of 6G-AKA, 5G-AKA is found to be vulnerable to several attacks, including linkability attack, DDoS attack, single-point-of-failure problem, and forward/post-compromise secrecy. 6G-AKA is expected to strengthen authentication between home network (HN) and serving network (SN) in order to prevent attacks which have been mentioned earlier, enable direct device-to-device authentication, bridge the gaps among heterogenous devices for their incompatible security capabilities, and revise its design for new subscriber identifier privacy model [4]. Besides, as DDoS attacks are becoming more complicated and threatening, it is crucial to make 6G-AKA more robust and equipped with security mechanisms to defend against DDoS attacks.

2) **Extensible Authentication Protocol – Transport Layer Security Plus (EAP-TLS+)**

EAP serves as an authentication framework used to support various authentication methods. TLS is one of the methods, which enables certificate-based mutual authentication within a private network. EAP-TLS is regarded as one promising authentication protocol and has been included in the annex of 5G security standard (TS 33.501). However, due to the size of current certificates (such as X.509v3), transmission of such certificates for authentication may cause too much overhead for infrequent data transmission in IoT scenarios. Thus, EAP-TLS+ is expected to support certificateless authentication methods, such as identity-based signature (IBS), implicit certificate. The IBS-supported EAP-TLS+ is able to establish device-to-device (D2D) mutual connection without the need of certificates. This feature will greatly benefit and facilitate vehicle-to-vehicle (V2V) and D2D wireless communications. In 6G, EAP-TLS+ is also expected to be further developed from 5G EAP-TLS and incorporate more advanced features, so as to offer a secure and seamless communication experience.

*C. Trustworthy Underlay*

Trustworthy undelay is indispensable for providing necessary support to higher level applications. Two types of underlay, covering decentralized authentication and network virtualization, are discussed as follows.

1) **Decentralized PKI (DPKI)**

PKI serves as an underlying framework which enables data encryption, digital signature creation and certificate-based authentication, propelling establishment of trust among involved entities. However, PKI in 5G is deployed in a centralized manner, which suffers from single point of failure.

PKI in 6G is expected to become more decentralized, which can be realized by blockchain, to avoid single point of failure caused by excessive dependence on a single root CA [9] and minimize control of third parties. The integration of blockchain will enable transparency and immutability in PKI, which means that certificate issuing can be observed by all entities, and issued certificates are traceable. In decentralized PKI, trust is decentralized, and built on consensus protocols. Developing an effective and decentralized PKI is critical for establishing a trustworthy identity management process in 6G networks.

2) **Network Function Virtualization/ Network Function Virtualization Infrastructure (NFV/NFVI)**

Owning to the emergence of numerous heterogenous devices as well as countless novel applications and various scenarios which require high flexibility and low latency, 6G is expected to incorporate NFV as an enabling technology for network virtualization. It is supported by NFVI, which consists of components of networking, computing and storage, and serves as a platform for VNFs. NFV enables services of different types to be run on top of commonly shared hardware appliances by decoupling network functions from proprietary hardware, and empowering a service-orientated networking. This feature will significantly increase network flexibility, scalability and reduce cost of network deployment [10]. Full virtualized 6G networks are expected to expedite harmonization, with core network, radio access networks and network edge underpinned by uniform underlying hardware. Furthermore, as AI pervades every corner of 6G networks, NFV can partner with AI to facilitate network automation, optimize resource utilization, and enhance quality of service (QoS).

## V. XENOGENESIS TRUST

In this section, we list the prospective technologies that can be used to undergird xenogenesis trust.

### A. Trusted Foundation

Cryptography forms the foundation of trustworthy communication. Common apply cryptographic techniques include symmetric cryptography and asymmetric cryptography. Given that symmetric cryptography decrypts and encrypts messages with the same key (which needs to be pre-shared before communication), it involves complicated key management problem. On the contrary, asymmetric cryptography doesn't need to share the key, and can enable direct authentication and secure communication among devices without the help of the core network device (such as home subscriber server (HSS) in 4G, unified data management (UDM) in 5G). Hence, cryptography in 6G is expected to gradually move from symmetric encryption to asymmetric encryption, serving as a strong foundation of xenogenesis trust.

On the other hand, with the powerful computing ability of quantum computers, the time needed to decrypt a key can be significantly reduced. Due to the security threats posed by quantum computing in future 6G network, a transformation is being fostered: from traditional cryptography to post-quantum cryptography [11]. The upgraded version will benefit asymmetric cryptography-based technologies, by preventing them from being compromised by quantum attacks.

### B. Trusted Platform

Trusted platform in this subsection mainly embodies X-Trust in terms of confidential computing and reliable data storage.

1) **Trusted Execution Environment (TEE)**

TEE, as a tamper-proof environment, is targeted to preserve code authenticity and data integrity within a device. It serves as an isolated secure zone in the main processors, in which unauthorized data access and malicious modification are prevented [12]. TEE is a key enabler for confidential computing. Current encryption methods mainly focus on ensuring integrity of data in storage and data in transmit, most of which do not focus on preserving integrity of data in use. Confidential computing, empowered by TEE, is targeted at preserving both data-in-use integrity and code confidentiality on device, facilitating establishment of end-to-end trustworthiness. Existing TEE solutions include ARM TrustZone, Intel SGX, as well as AMD Secure Encrypted Virtualization (SEV). Besides, TEE can be leveraged to ensure data privacy of cloud computing. As encrypted data needs to be decrypted in cloud services to facilitate data processing, data in use becomes vulnerable. TEE will offer a trustworthy execution environment such that data privacy is preserved even the data is being handled by third parties. It is also envisioned to be meld with network functions

2) **Blockchain**

Blockchain is a distributed ledger characterized by being highly immutable. Blockchain is structured by a chain of blocks, a new block will be added to which only after being verified through a consensus mechanism. Since the chain is one-directional, an operation on blockchain is irreversible, and the recorded data cannot be modified. The irreversible nature of blockchain will facilitate establishment of mutual trust between different parties and enhancement to privacy preservation. For instance, as spectrum sharing, automated orchestration, decentralized computation will be widely applied in 6G networks, blockchain will be able to facilitate distribution and management of resources in a secure and privacy-preserving manner. Beyond this, it can also be embedded in authentication and authorization process for key management and access control. Application of blockchain in 6G networks will increase the security level, in terms of privacy, data integrity and service availability, and enable massive connectivity with assurance of trustworthiness [13].

### C. Trusted Hardware

In this subsection, we present a significant component of X-Trust-the trusted hardware, including TPM and various SIMs.

1) **Tamper Proof Module (TPM)**

TPM is a tamper-proof module dedicatedly designed to establish hardware root of trust (RoT) by securing hardware and securely storing encryption keys, certificates, or other confidential information for platform authentication. In a highly distributed and virtualized network environment of 6G era, hardware RoT becomes critical for assuring communication security especially on untrusted platforms. TPM is envisioned to empower trusted computing in NFV in two ways: integrity preservation with secure storage, as well as trustworthy verification with remote attestation [14]. During the boot process, measurement values of system components will be sheltered and cannot be modified during run time. Remote attestation is then applied to remotely verify whether the system's booting process can be trusted given its measurement values at the load time. Furthermore, recognizing the increasingly growing number of heterogenous devices on the edge, deployment of trustworthy TPM is able to enhance their tamper resistance and efficiently reduce the vulnerability.

2) **eSIM/iSIM**

In 6G, reliable and trustworthy massive machine-to-machine (M2M) communications are of much importance. A traditional approach is SIM, a removable smartcard used for subscriber identification and authentication. However, as a physical object, it needs to be plugged in and out for every IoT device, which makes it troublesome to be deployed in massive IoT networks. To overcome the limitations of traditional SIM, embedded SIM is subsequently proposed and deployed [15]. Compared to removable SIM, eSIM enables operator profiles to be provisioned "over the air". This underlying feature offers a seamless communication process for heterogenous devices deployed around the world. Apart from eSIM, another type of SIM is integrated SIM (iSIM), supported by system-on-chip (SoC). eSIM and iSIM are predicted to become enabling technologies in many 6G IoT verticals, including but not limited to smart factories, eHealth, smart grids and connected autonomous vehicles, and will facilitate cellular M2M communication, secure updates of firmware, and ensure flexible and trusted IoT connectivity [15].

3) **eTPSIM**

As mentioned previously, traditional SIMs may be no longer suitable to massive devices. TPM, as a trustworthy and secure chip, is envisioned to be integrated with SIM to in order to develop embedded TPSIM (eTPSIM), especially for mobile devices. The integration provides a unified solution to device identity authentication, trust booting, and platform integrity. eTPSIM can be soldered into a device's circuit board, which facilitates the establishment of physical binding between root of trust for measurement of the platform and eTPSIM. It will become beneficial for many large-scale 6G applications: Internet of Vehicles (IoV), smart cities, Industrial Internet of Things (IIoT), etc. It can effectively reduce cost of SIM deployment among numerous mobile devices, enhance security and trustworthiness of IoT terminals, empower trusted computing for IoT devices, and assure information security for critical information infrastructure.

## VI. Conclusion

In this article, we have proposed a novel trust framework SIX-Trust for building a more secure and trustworthy 6G networks. The framework consists of three layers: sustainable trust layer, infrastructure trust layer, and xenogenesis trust layer. Each layer focuses on a different aspect of trust for 6G. For each layer, we have given our insights on the potential technologies can be used, how the technologies can facilitate establishment of trustworthiness in 6G, and why they are crucial for future trustworthy 6G networks.

## Biographies

**Yiying Wang** (ywang109@e.ntu.edu.sg) is currently pursuing the B.S. degree in computer science with Nanyang Technological University, Singapore. The article is accomplished during her internship which mainly focuses on 6G and trust modeling at the Digital Identity and Trustworthiness Laboratory, Huawei Singapore. Her research interests include artificial intelligence and digital security and plans to pursue further studies in the areas.

**Xin Kang** (kang.xin@huawei.com) is senior researcher at Huawei Singapore Research Center. Dr. Kang received his Ph.D. Degree from National University of Singapore. He has more than 15 years' research experience in wireless communication and network security. He is the key contributor to Huawei's white paper series on 5G security. He has published 70+ IEEE top journal and conference papers, and received the Best Paper Award from IEEE ICC 2017, and Best 50 Papers Award from IEEE GlobeCom 2014. He has also filed 60+ patents on security protocol designs, and contributed 30+ technical proposals to 3GPP SA3. He is also the initiator and chief editor for ITU-T standard X.1365, X.1353, and the on-going work item Y.atem-tn.

**Tieyan Li** (Li.Tieyan@huawei.com) is currently leading Digital Trust research, on building the trust infrastructure for future digital world, and previously on mobile security, IoT security, and AI security at Shield Lab., Singapore Research Center, Huawei Technologies. Dr. Li is also the director of Trustworthy AI C-TMG and the vice-chairman of ETSI ISG SAI. Dr. Li received his Ph.D. Degree in Computer Science from National University of Singapore. He has more than 20 years experiences and is proficient in security design, architect, innovation and practical development. He was also active in academic security fields with tens of publications and patents. Dr. Li has served as the PC members for many security conferences, and is an influential speaker in industrial security forums.

**Haiguang Wang** (wang.haiguang.shieldlab@huawei.com) is a senior researcher on identity, trust and network security in Huawei International Pte. Ltd. He received a Ph.D. degree in Computer Engineering from National University of Singapore in 2009 and a Bachelor from Peking University in 1996. He joined Huawei at year 2013 and currently he is a senior researcher at Huawei. He was a research engineer/scientist at I2R Singapore since 2001. He is an IEEE Senior member.

**Cheng-Kang Chu** (Chu.Cheng.Kang@huawei.com) received his Ph.D. in Computer Science from National Chiao Tung University, Taiwan. He is a senior researcher of Huawei International, Singapore. Dr. Chu has had a long-term interest in the development of new technologies in applied cryptography, cloud computing security and IoT security. His research now focuses on mobile security, IoT security, decentralized digital identity, Web 3.0, etc. Dr. Chu has published many research papers in major conferences and journals like PKC, CT-RSA, AsiaCCS, IEEE TPDS, IEEE TIFS, etc. and received the best student paper award in ISC 2007.

**Zhongding Lei** (lei.zhongding@huawei.com) is a senior researcher at Huawei Singapore Research Center. He has been working on 5G network security since 2016. Prior to joining Huawei, he was a laboratory head and senior scientist with the Agency for Science, Technology, and Research (A-STAR) of Singapore, involved in research and development of 3GPP and IEEE standards in wireless systems and networks. He has been the Editor-in-Chief of IEEE Communications Standards Magazine since 2019.